# Inflight performance of the GECAM Gamma-ray and Charge particle Detectors

X.Q. Li[1*], X.Y. Wen[1], S.L. Xiong[1], K. Gong[1], D.L. Zhang[1], Z.H. An[1], Y.B. Xu[1], Y.Q. Liu[1], C. Cai[1], Z. Chang[1], G. Chen[1], C. Chen[1], Y.Y. Du[1], M. Gao[1], R. Gao[1], D.Y. Guo[1], J.J. He[1], D.J. Hou[1], Y.G. Li[1], C. Li[2], C.Y. Li[3], G. Li[1], L. Li[1], Q.X. Li[2], X.F. Li[1], M.S. Li[1], X.H. Liang[1], X.J. Liu[1], F.J. Lu[1], H. Lu[1], X. Ma[1], B. Meng[1], W.X. Peng[1], R. Qiao[1], F. Shi[1], L.M. Song[1], X.L. Sun[1], H. Wang[1], H.Z. Wang[1], J.Z. Wang[1], P. Wang[1], Y.S. Wang[1], X. Wen[1], S. Xiao[1], Y.P. Xu[1], S. Yang[1], J.W. Yang[1], Q.B. Yi[2], Fan. Zhang[1], S.N. Zhang[1], C.Y. Zhang[4], C.M. Zhang[1], Fei Zhang[1], K. Zhang[5]，P. Zhang[6], X.Y. Zhao[1], Y. Zhao[7], S.J. Zheng[1], X. Zhou[1]

[1] *Institute of High Energy Physics, CAS, Beijing 100049, China*
[2] *Xiangtan University, Xiangtan 411105, China*
[3] *China West Normal University, Nanchong 637002, China*
[4] *Changchun University of Science and Technology, Changchun 130022, China*
[5] *Qufu Normal University, Qufu 273165, China*
[6] *Southwest Jiaotong University, Chengdu 611756, China*
[7] *Beijing Normal University, Beijing 110875, China*
* E-mail: lixq@ihep.ac.cn

**Abstract** The GECAM mission consists of two identical microsatellites (GECAM-A and GECAM-B). Each satellite is equipped with 25 gamma-ray detectors (GRD) and 8 charged particle detectors (CPD). The main scientific objective of the GECAM mission is to detect gamma-ray bursts (GRBs) associated with the gravitational wave events produced by the merging of binary compact stars. After the launch on Dec. 10, 2020 , we carried out a series of on orbit tests. This paper introduces the test results of the GECAM-B satellite. According to the in-flight performance, the energy band for gamma-ray detection of GECAM-B is from about 7 keV to 3.5 MeV. GECAM-B can achieve prompt localization of GRBs. For the first time, GECAM-B   realized a quasi-real-time transmission of trigger information using Beidou-3 RDSS.

## 1. Introduction

The observation of GW170817 is the first time that human beings use gravitational waves and electromagnetic waves to observe the physical process of the same object. It is also the first and only time that humans observe gravitational wave and its electromagnetic wave, indicating the arrival of the era of multi messenger gravitational wave astronomy[1-8]. The GECAM (Gravitational wave high-energy Electromagnetic Counterpart All-sky Monitor) satellite(also named "Huairou-1") aims to detect gamma-ray bursts, which is related to the gravitational waves of double compact object mergers[1,9,10]. The GECAM satellite consists of two identical satellites developed by the Chinese Academy of Sciences. GECAM can detect and investigate various outbursts of high-energy celestial sources[1]. The Institute of high energy physics (IHEP) is the project proposer and payload developer. GECAM carries two types of detectors. The first detector is the gamma-ray detector (GRD), which realizes the spectral and temporal detection in the 7 keV ~ 3.5 MeV energy band; the other is the charged particle detector (CPD), which can detect energetic electrons in the 260 keV ~ 6.6 MeV. Each satellite is equipped with 25 GRDs and 8 CPDs. These detectors are nearly evenly arranged along the hemispheric cupola on the satellite dome. Using a fitting approach, GECAM can use the counts rate distribution of each detector to calculate the direction of a gamma-ray burst (GRB), i.e. localization [9,10].

The GECAM satellite was launched on December 10, 2020 (Beijing Time). These two satellites entered the predetermined round orbit with an inclination of 29 degrees and a height of 600 km. The field of view of a single satellite is the all sky unblocked by the Earth. . Since launch, only the payload of GECAM-B has been working. The payload of the GECAM-A satellite has not been turned on yet due to the power supply issue. The inflight test of the payload of GECAM-B was carried out for six months (From 15 Jan., 2021 to 15 Jun., 2021).

## 2. Design of the payload

Li et al. [9,10] described detailed design characteristics of the payload of GECAM. The basic parameters of the two kinds of detectors (GRD and CPD) and the electronics box (EBOX) of GECAM are shown in Table 1, where a summary of the design of GECAM payload is provided, for completeness.

Table 1. The basic parameters of GECAM (single satellite index) [9,10]

| Detector | Item | Requirement | Value |
|---|---|---|---|
| GRD | Number of detectors | ≥20 | 25 |
| | Shell (housing?) | Aluminum | Aluminum |
| | Window | Beryllium + ESR | Beryllium + ESR |
| | Window thickness | ≤300μm Be | 200μm Be + 64μm ESR |
| | Sensitive material | Lanthanum bromide crystal | Lanthanum Bromide crystal |
| | Crystal thickness | ≥15mm | 15mm |
| | Crystal diameter | ≥76mm | 76.2mm |
| | Photoconductive | Glass | K9 Glass |

|  | material | | |
|---|---|---|---|
| | Photoelectric converter | SiPM | SiPM (MicroFJ-60035-TVS) |
| | Number of photoelectric converters per detector | - | 64 |
| CPD | Number of detectors | ≥6 | 8 |
| | Shell | Aluminum | Aluminum |
| | Window | Aluminized polyimide film | Aluminized polyimide film |
| | Window thickness | ≤20μm Polyimide+≤200μm Teyek | 13μm Polyimide +5nm Aluminum+162μm Teyek |
| | Sensitive material | Plastic scintillator BC408 | Plastic scintillator BC408 |
| | Plastic scintillator size | 40mm*40mm*10mm | 40mm*40mm*10mm |
| | Calibration of radioactive source | $^{241}$Am | $^{241}$Am |
| | Activity of radioactive source | 10 ~ 50 Bq | 20 ~ 30 Bq |
| | Photoelectric converter | SiPM | SiPM (MicroFJ-60035-TVS) |
| | Number of photoelectric converters per detector | - | 36 |
| Electronics box | Number of data acquisition channels | Depending on the number of probe signal channels | GRD: high gain: 25 low gain: 25 CPD: 8 |
| | Primary power supply voltage | 28 V | 28 V |
| | Data transmission protocol | RS-422 | RS-422 |
| | Bus protocol | CAN | CAN |
| | Redundant design | Avoid a single point | Data acquisition circuit: Each channel serves as backup for each other. Data management circuit: cold backup. Power supply: cold backup. |

The payload of each of the GECAM satellites consists of 25 gamma ray detector (GRD), 8 charged particle detector (CPD) and an electronics box (EBOX). All detectors have modular designs[9-11].

As shown in Figure 1, each satellite has 25 GRD modules, which are mounted on the dome of the satellite. Each module can detect the gamma ray photons from the front-facing region, and the field of view is ~2 π. 25 GRDs point to different directions. On-orbit localization of transient sources, including gamma-ray bursts, magnetar bursts, solar flares, was conducted with a mesh-based look-up-table fitting approach [9,10,12].

The GRD adopts the technology of lanthanum bromide crystal readout with SiPM[13]. The basic parameters of the GRD are shown in Table 1.

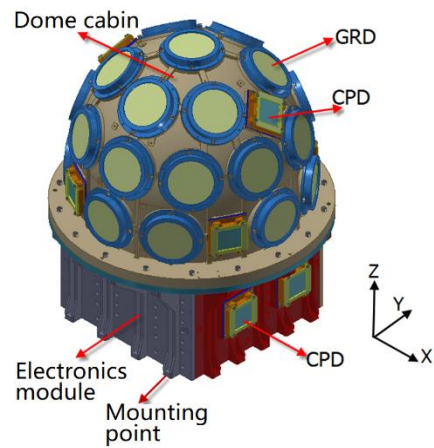

Figure 1 The payload of GECAM[10]

Each satellite has 8 CPD modules, which are in different directions of the satellite surface (except the direction towards the earth). The CPD needs to monitor the flux of charged particles on the GECAM orbit and help to distinguish between gamma ray bursts and space charged particle events. The CPD adopts the technology of matching plastic scintillator with SiPM[9,10]. The basic parameters of the CPD are shown in Table 1.

The EBOX is in the payload electronics module, which is used to collect and process GRD and CPD data to realize the on orbit triggering and positioning of gamma ray bursts, realize data storage, packaging, and transmission, as well as interaction with satellite platform, and provide a power supply. The main structure is used for the installation and arrangement of GRD and CPD detectors, and docking with the satellite platform[9,10].

### 3. Ground-based performance test

The design and testing of the GECAM will be published in the fourth issue of Radiation Detection Technology and Methods (RDTM, https://www.springer.com/journal/41605) in 2021 as a collection of papers. All probes of the GECAM have gone through PCB level stress screening and a component quality assurance test. The GRD test probes and CPD flight probes have been tested for thermal vacuum and mechanical properties. EBOX has carried out the stress screening test, thermal vacuum test, atmospheric thermal cycle test and mechanical test. After the whole machine is assembled, the EMC test and magnetic test has been carried out for the GECAM payload.

The calibration test for the two kinds of probes on the GECAM has been carried out for more than half a year. The calibration items of the GRD and CPD include:

- Energy response: E-C relationship (γ for GRD and $e^-$/p for CPD)/ Energy resolution (γ for GRD and $e^-$/p for CPD)/response matrix file (for the GRD).
- Detection efficiency (γ & $e^-$, both)/the influence of coating multilayer on detection efficiency (γ for GRD and $e^-$ for CPD).
- Spatial response: Influence of X/γ-ray incident position (GRD) & direction (γ for GRD and $e^-$ for CPD).
- Temperature and bias voltage response (γ for GRD and $e^-$ for CPD).
- Dead time (γ for GRD and $e^-$/p for CPD).
- Response to energetic electrons (GRD).

In addition, we have carried out the following special tests according to the characteristics of GECAM:
- Large signal response test
- Response test under high count rate
- Trigger and positioning test
- Time delay test of two GECAM satellites
- Relative time accuracy
- Air tightness test of GRD

## 4. Launch and in-flight performance

The GECAM satellites were launched into orbit via a CZ-11 solid-fueled carrier rocket at Xichang Satellite Launch Center at 04AM on December 10, 2020 (Beijing Time). After the separation of the satellite and rocket, the solar panels of the two GECAM satellites unfolded normally. Each GECAM satellite has two sets of solar panels. Due to the power supply issues, the GECAM-02 satellite payload has power to work for about half an hour each day, while the GECAM-01 satellite has still been in shortage of power supply thus unable to turn on the payload.

On January 14, 2021, the GECAM-02 satellite was adjusted to the observation mode with the solar panel normal direction pointing to the sun to maximize its energy output. Since then, the payload of the GECAM-02 satellite has been available for observation for about 11 hours per day. The in-flight test of GECAM-02 satellite payload mainly includes:

1) Working condition inspections and basic working parameter settings of the payload.

2) Function and performance tests of the payload:
- The working parameters, on-orbit trigger and positioning parameters, temperature-bias voltage adjustment parameters optimization and update are determined.
- time system test, data storage and downlink transmission mode test, trigger

information transmission link (using Beidou navigation system) test.
- South Atlantic Anomaly (SAA) region adjustment, observation direction of sky area optimization, power on-off strategy optimization, temperature control strategy optimization.

3) On-orbit calibration.

This paper summarizes in-flight measurements and associated analyses for some of the most important performance characteristics of GECAM, including:

1) Spectrum response: energy range, energy resolution, background.
2）Dead time.
3) Sensitivity.
4) Positioning accuracy.
5) On orbit calibration.

**4.1 Spectrum response**

Both the GRD and CPD detectors record the energy and time information of each physical event by case detection. The counting rate of endogenous background produced by lanthanum bromide crystal of GRD is about 100 counts/s [13]. The activity of the Am-241 radiation source implanted into CPD is 20-30 Bq. In addition to the background signal inside the probes, the background energy spectrum during in-orbit operation also includes diffuse X-ray, space charged particle, high-energy celestial body source background, atmospheric reflection X-ray, cosmic rays and secondary particles generated by their interaction with satellite materials [13].

For the energy spectrum analysis of high-energy celestial explosion events, such as gamma ray bursts, the background energy spectrum of each probe needs to be deducted. The spectral lines in the background spectrum can be used for on orbit calibration. Both GRD and CPD can adjust the gain by updating the bias voltage table (bias voltage - temperature relationship of each probe). The two kinds of detectors can adjust the bias voltage value in real time according to the temperature, to keep the gain stable[14,15,16].

Figure 2 shows the comparison between the background energy spectrum measured by the two detectors at room temperature of the launch site before satellite launch and the background energy spectrum measured at - 20 ° C on orbit. Compared with the ground local energy spectrum, 86keV, 190keV and 511keV spectral lines are added to the on-orbit background energy spectrum of GRD[14]. These spectral lines come from on orbit particle excitation and positron annihilation lines. By comparing the peak positions of endogenous spectral lines, the gain on the ground with a difference of 45 ℃ is consistent with that on orbit, which indicates that the automatic gain control function is normal.

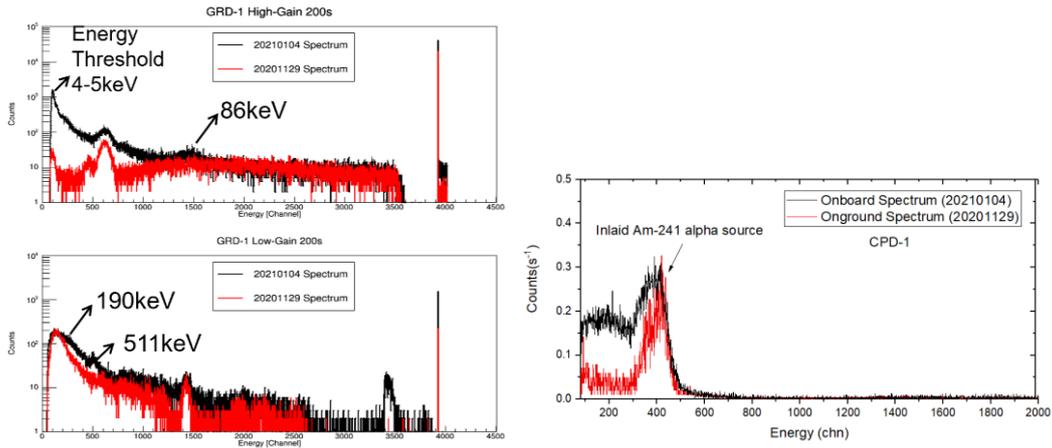

Figure 2. Comparison between ground energy spectrum and on orbit energy spectrum. The red line in the figure is the ground energy spectrum and the black line is the on-orbit energy spectrum. The left figure shows the GRD results, and the right figure shows the CPD results.

By analyzing the on-orbit energy spectrum of each probe, we get their energy range and energy resolution, as shown in Table 2 and Table 3.

In Table 2, the energy range of each probe of GRD covers 7.1keV-3.57MeV, satisfying the scientific requirements of 8keV – 2MeV. The energy resolution is 10.1% - 16.1% @ 59.5keV, meeting the technical index requirements of < 18% @ 59.5keV[16]. To facilitate the on-orbit trigger positioning calculation, the gain consistency of each GRD probe is achieved by adjusting the bias voltage. The consistency result of the adjusted probe gain is shown in Figure 3. Figure 3 shows the consistency of each probe, with 37.4keV of high gain and 1.47MeV of low gain. GRD uses two doped lanthanum bromide crystals: two-component doping of 5% Ce + 1.5% Sr and one-component doping of 5% Ce. Figure 3 shows that even though the deviation of individual double doped probes reaches - 3.71%, the consistency is generally better than ± 2.2%[9,15,16].

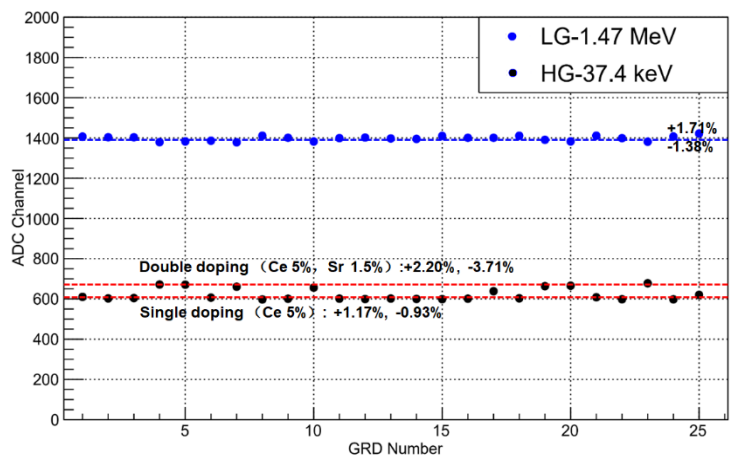

Figure 3. Consistency of peak position of GRD of GECAM 02 satellite. The data points on two upper and lower red lines is corresponded to high gain 37.4keV peak

position of GRD with double doped and single doped crystals; the blue data points refer to low gain 1.47MeV peak position[9]

In Table 3, the energy range of CPD probes covers 182keV - 6.61MeV (CPD 1 - 6) and 263keV – 8.11MeV (CPD 7/8), meeting the scientific requirements of 300keV – 5MeV. The energy resolution is 26.5% - 37.9% @ 570keV (there is no spectral line analysis for space particle detection, so there is no technical index requirement for energy resolution). CPD07 and CPD08 have higher energy ranges than other CPD probes because they are located on the side plate of the electronics cabinet and have higher temperatures [17].

Table 2. On orbit energy range and resolution of each GRD of GECAM-02 satellite

| GRD number | Energy range(keV) | Energy resolution%(@37.4keV) | Equivalent energy resolution %(@59.5keV) |
| --- | --- | --- | --- |
| 01 | 4.1-3576 | 17.32 | 13.92 |
| 02 | 5.3-4369 | 16.34 | 13.13 |
| 03 | 5.5-4443 | 15.59 | 12.53 |
| 04 | 6.6-5627 | 15.38 | 12.36 |
| 05 | 6.8-5706 | 14.48 | 11.63 |
| 06 | 5.8-4703 | 14.81 | 11.90 |
| 07 | 5.8-5180 | 16.36 | 13.15 |
| 08 | 5.9-4518 | 16.40 | 13.18 |
| 09 | 4.3-3890 | 19.10 | 15.35 |
| 10 | 6.9-5680 | 14.74 | 11.84 |
| 11 | 4.4-3606 | 13.20 | 10.61 |
| 12 | 6.4-4785 | 13.58 | 10.91 |
| 13 | 6.3-4917 | 15.00 | 12.05 |
| 14 | 4.8-4809 | 14.76 | 11.86 |
| 15 | 5.4-4464 | 20.08 | 16.13 |
| 16 | 4.6-3573 | 15.13 | 12.16 |
| 17 | 4.8-3702 | 15.25 | 12.25 |
| 18 | 5.3-4218 | 16.31 | 13.11 |
| 19 | 7.1-6225 | 17.93 | 14.41 |
| 20 | 6.8-5982 | 12.59 | 10.12 |
| 21 | 6.4-4924 | 15.88 | 12.76 |
| 22 | 6.7-5346 | 15.84 | 12.73 |
| 23 | 7.0-5553 | 14.35 | 11.53 |
| 24 | 6.4-5434 | 14.93 | 12.00 |
| 25 | 5.2-4165 | 17.39 | 13.97 |

Table 3. On orbit energy range and resolution of each CPD of GECAM-02 satellite.

| CPD number | Lower energy threshold (keV) | Upper energy threshold (keV) | Energy resolution% @ α-peak* of Am-241 |
|---|---|---|---|
| 01 | 157.65 | 6609.96 | 28.5 |
| 02 | 168.15 | 7039.17 | 31.7 |
| 03 | 177.27 | 7428.00 | 30.4 |
| 04 | 164.96 | 6910.91 | 28.3 |
| 05 | 167.34 | 7029.30 | 26.5 |
| 06 | 181.92 | 7135.13 | 37.9 |
| 07 | 257.91 | 8110.90 | 33.3 |
| 08 | 262.55 | 8434.37 | 37.2 |

*: 5.486MeV, equivalent electron energy: 570keV

### 4.2 Dead time

The dead times for normal events of GRD and CPD are designed to be 4 $\mu$s and 4.8 $\mu$s respectively. To verify the on orbit dead time of these two kinds of detectors, we use on orbit data analysis to obtain their time interval spectrum, as shown in Figure 4. It can be seen from the figure that the time interval spectra obey exponential distribution, and the minimum time interval (i.e. dead time) of the two kinds of detectors is consistent with the design.

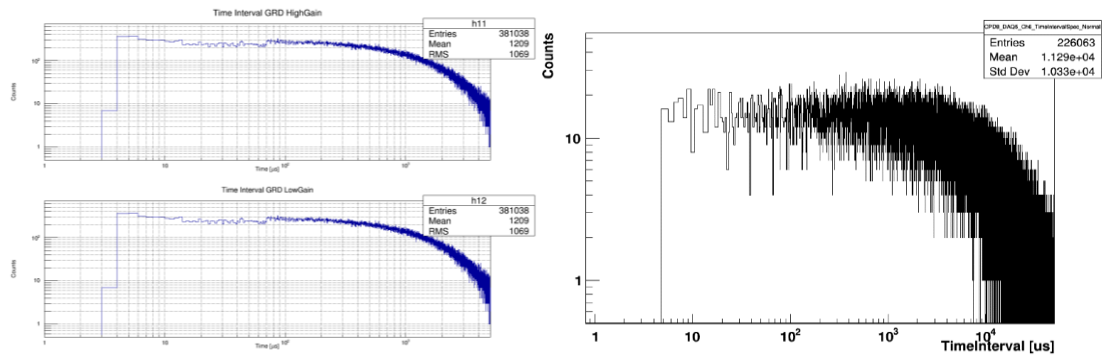

Figure 4. Time interval spectrum of GRD and CPD. Left: GRD high gain and GRD low gain, right: CPD.

### 4.3 Sensitivity

A section of on orbit background data close to 700 counts/s is randomly selected, and the average background of a single probe is about 685 counts/s, as shown in Figure 5.

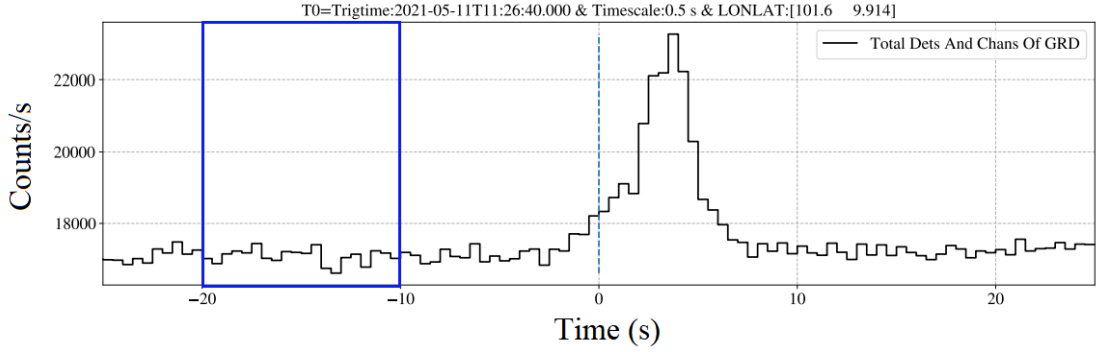

Figure 5. The randomly selected background data (the blue box) for sensitivity calculation.

Using the on-orbit calibration results, an arbitrary incident direction (theta = 2.92 °, phi = 45 °) is selected to produce the instrument response and the expected detector deposition count. We analyze the data as follows:

1. Background counting rate using the data of the selected time period on the satellite: ~685 counts/s/GRD.

2. Input soft X-ray band spectrum to obtain the expected counting rate of instrument response in a certain direction.

3. If the detected net count and expected count are consistent in the relative ratio of each probe and energy channel, the best weight factor will be obtained, so it is assumed that they are only different in intensity (controlled by Amplitude).

4. Obtain the value of Amplitude (A) according to the index of 3 sigma.

5. Calculate the flux according to the energy spectrum and A to obtain the sensitivity.

For the selected data, the sensitivity curve obtained according to the steps 1-5 is shown in Figure 6.

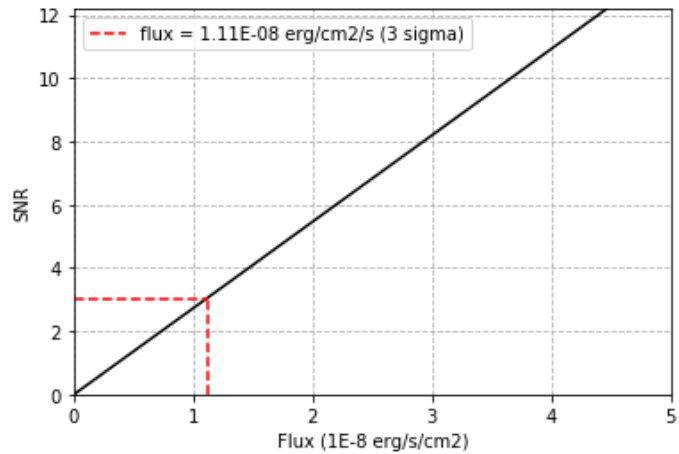

Figure 6 Detection sensitivity curve of GECAM02 satellite (700 counts/s background level and soft band spectrum duration of 20s).

According to the results in Figure 6, the gamma ray burst detection sensitivity corresponding to the optimal weighted 3 $\sigma$ signal-to-noise ratio is 1.11e-08

erg/cm$^2$/s (10 keV-1000 keV, 685cps background, band soft spectrum, 20s signal duration), which meets and is better than the index requirements of 2e-08 erg/cm$^2$/s (under the condition of 3 Sigma signal-to-noise ratio, 700cps background level and soft band spectrum duration of 20s).

### 4.4 Positioning accuracy

The GECAM positioning error index refers to the comprehensive positioning error of double satellites, but at present, only GECAM-02 is in observation operation. Next, we analyze the positioning error of single GECAM satellite, and then estimate the positioning error of the double satellites according to the on-orbit observation.

We select the gamma ray bursts with the best energy spectrum information measured by the GECAM-02 satellite in each stage of on orbit observation, as shown in Table 4.

Table 4. Selected GRB events information

| GRB number | GECAM trigger time (UTC) | GECAM On board positioning error (1-sigma) | Fluence (erg/cm2, 10-1000 keV) | Cumulative time (s) | Normalized positioning error (1-sigma) | Incident angle in the payload coordinate system as shown in Figure 1(theta, phi), deg |
|---|---|---|---|---|---|---|
| GRB 210119A | 2021-01-19T02:54:09.850 | 4.2 | (3.4+/-0.4) E-7 | 0.05 | 4.16 | (10.58, 44.84) |
| GRB 210204A | 2021-02-04T06:30:00.600 | 3.33 | (5.76+/-0.04)E-6 | 53 | 2.85 | (117.32, 198.74) |
| GRB 210207B | 2021-02-07T21:52:14.050 | 1 | (7.6+/-0.3)E-6 | 106 | 0.26 | (122.78, 46.58) |
| GRB 210511B | 2021-05-11T11:26:40.600 | 3.19 | (1.04+/-0.02)E-5 | 8 | 3.17 | (121.62, 37.12) |

The positioning capability of the satellite is affected by background level, GRB brightness, spectrum shape, duration, incidence angle of the payload coordinate system and other factors. The simulation results of GECAM show that the positioning accuracy of a medium hardness spectrum with typical brightness ($10^{-6}$ erg s$^{-1}$ cm$^{-2}$ for 10 s at 10 – 1000 keV energy band) is about 0.6° (1-σ Statistical error), and the positioning accuracy is correlated with the logarithm of flow and duration (LIAO Jin-Yuan, et al.).

According to the simulation results, the positioning accuracy and duration are approximately in line with a piecewise log linear relationship in the range of 0.01 ~ 10 s and 10 ~ 100 s. The linear coefficient in the range of 0.01 ~ 10 s is about 0.1 and the linear coefficient in the range of 10 ~ 100 s is about 1.0 (LIAO Jin-Yuan, et al.). According to these coefficients, the different GRB samples in Table 4 are normalized to the positioning accuracy when the flux is $10^{-6}$ erg s$^{-1}$ cm$^{-2}$ and lasts for 10 s. (see the normalized positioning error in Table 4). Thus, the average on-board positioning accuracy of GECAM-02 for medium brightness gamma bursts (fluence is $10^{-6}$ erg cm$^{-2}$ and duration is 10 s) is approximately 2.61 degrees.

To estimate the positioning error of the pair of GECAM satellites, we use Fermi / GBM (Charles Meegan, et al., 2009) as the approximate representative of GECAM-01 to conduct a joint time-delay positioning with GECAM-02, as shown in Figure 7.

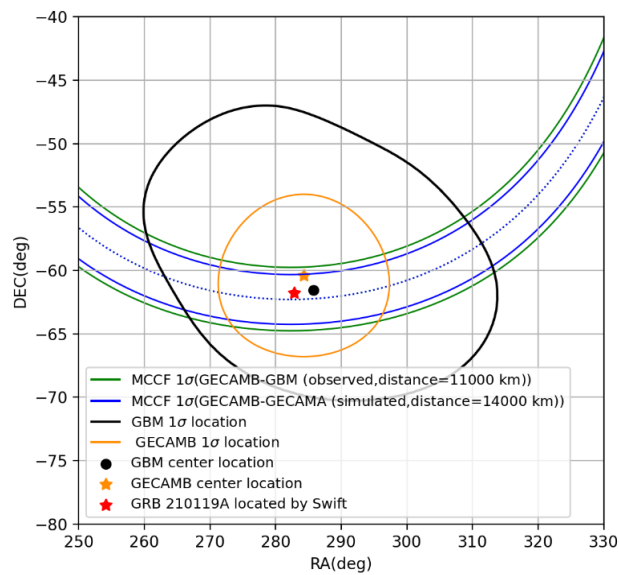

Figure 7. Positioning of GRB 210119A

In Figure 7, the time-delay positioning results of GRB 210119A are shown using a green line. The actual distance between GECAM-02 and Fermi is 11000 km, and the half width of positioning error is about 2.5 degrees. The result is converted to the distance (14000 km) between the two GECAM satellites during normal observation operation, and the positioning error of GECAM-A GECAM-B is 2.0 degrees (1-sigma). The fluence and duration of GRB 210119a is further converted to the positioning error for medium brightness gamma ray burst (fluence is $10^{-6}$ erg·cm$^{-2}$, duration is 10 s): 2.0 / SQRT (1e-5 / 3.4e-7) = 0.4 degrees.

Based on this analysis, which uses the gamma burst samples measured during the orbit operation of GECAM-B satellite, the on-board positioning accuracy of the GECAM-B single satellite is about 2.6 degrees (1-sigma, statistical error) for medium brightness gamma ray bursts (fluence is 10-6 erg·cm-2, duration is 10 s). If Fermi / GBM is used to replace the GECAM-01 satellite for equivalent conversion, the positioning error of the two GECAM satellites is estimated to be 0.4 degrees (1-sigma, statistical error). These results meet the ~1 degree design index requirements.

### 4.5 On orbit calibration

The Lanthanum bromide crystals carry several endogenous spectral lines, and those available for on-orbit calibration of the GRD were mainly the 37.4keV and 1.47MeV spectral lines. In addition, the other lines available for on-orbit calibration mainly include the 85.8keV on-orbit activation spectral line of Bromine in lanthanum bromide crystals and the 511keV positron electron annihilation line in the on-orbit background spectrum.

By accumulating the on-orbit data, the ADC spectrum of high and low gain of each GRD detector can be obtained. For high gain ADC spectra, the peak positions and sigma of the 37.4 keV and 85.8 keV peaks can be fitted. For low gain ADC spectra, the peak positions and sigma of 511 keV and 1470 keV peaks can be fitted as shown in Figure 8.

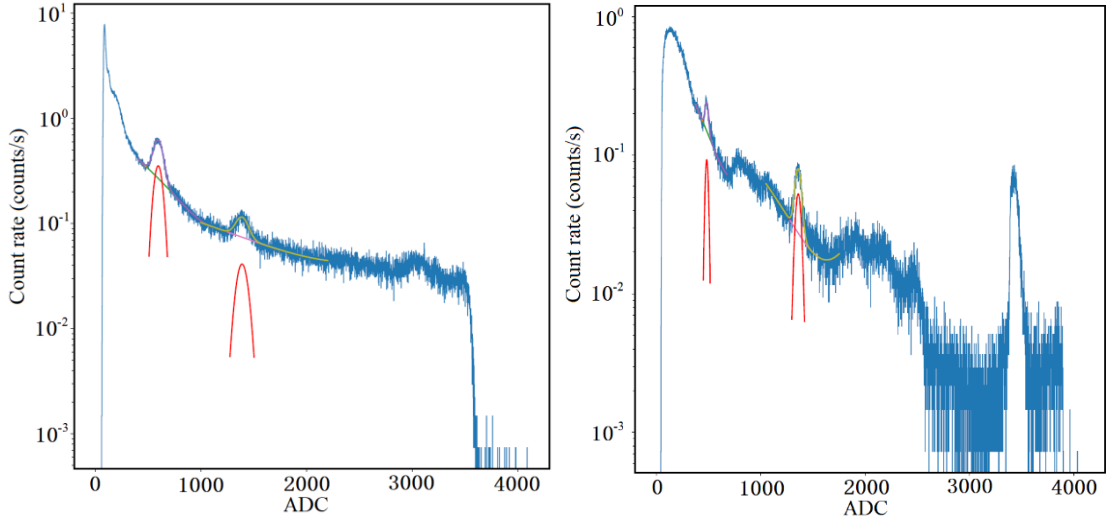

Figure 8. On-orbit background spectrum of GECAM-02 GRD 1. Left: High gain; the two Gaussian fitted peaks correspond to 37.4 keV and 85.8 keV; Right: Low gain; the two Gaussian fitted peaks correspond to 511 keV and 1470 keV.

The E-C relationship and energy resolution of each GRD can be obtained according to the Gaussian fitting results of each characteristic peak.

Considering that the CPD is only used to identify charged particle events, there were no requirements on the CPD's energy resolution. Since the energy spectrum of space charged particles has no spectral structure, 59.5keV characteristic line of CPD imbedded source $^{241}$Am can be used to roughly calibrate the change of gain.

## 5. Summary and discussion

An on-orbit test of GECAM satellite has been carried out. During the on-orbit test, we conducted a comprehensive test of various functions and performances of the payload. By fine tuning the bias voltage of each probe, we improved the gain consistency of each probe. By fitting the background characteristic spectrum, the on-orbit E-C relationship was obtained. At the same time, the energy resolution of each probe in orbit was verified. Through the analysis of on orbit data, the dead time, sensitivity, and positioning accuracy meet the design index requirements.

The inflight indexes of GECAM-B satellite are summarized in Table 5.

Table 5. The inflight indexes of GECAM-B satellite at the end of on orbit test.

| On orbit test object | Item | Performance index |
|---|---|---|
| The whole payload | sensitivity | 1.11E-08 erg/cm$^2$/s (3 Sigma SNR, ~ 700 counts/s background level and 20 s Band soft spectrum duration) |
| | positioning accuracy | 2.61°(Single star)/0.4°(Double stars，Taking Fermi GBM as equivalent to GECAM-01 satellite), Fluence:10$^{-6}$ erg·cm$^{-2}$，Duration: 10 s |
| GRD | Number of probes | 25 |
| | Energy range | 7.1keV～3.56 MeV |
| | Effective area | 45.3 cm$^2$ |
| | Dead time | 4 μs |
| | Energy resolution | ≤16.1%@59.5 keV |
| | detection efficiency | 78%@8keV |
| CPD | Number of probes | 8 |
| | Energy range (e-) | 258 keV～6.61 MeV |
| | Gamma ray detection efficiency | <7%@8-2000 keV |
| | Dead time | 4.8μs |
| Electronics box | Calculation time of on orbit trigger and positioning | <1s |
| | Relative time accuracy of each probe | 0.3μs |

The GECAM satellite payload uses lanthanum bromide crystal matching SiPM technology. The performance of each detector will change under the influence of temperature, irradiation, and other conditions. At present, the working state of payload has been stable. Subsequent on orbit calibrations will continue throughout the whole life of the payload.

On the other hand, there are approximately ten times the on-orbit triggering from various bursts every day since inflight operation. The GECAM satellite used a global RDSS communication link from the Beidou-3 navigation system (the first time this have ever been done), which can transmit the trigger positioning information of burst to the ground receiving station within about 1 minute. We believe that the Beidou-3 navigation system's global RDSS communication link will be widely used in subsequent astronomical satellites.

**Acknowledgements**

The authors wish to thank Hu Tai, Wu Haiyan, Zhang Wei, Bai Meng, Zhang Keke,

Huang Jia, Han Xingbo for their support during the satellite on orbit test. This project is supported by National Natural Science Foundation of China (12173038) and the strategic leading science and technology program (XDA 15360100，XDA 15360102) of the Chinese Academy of Sciences.